\documentclass[a4paper]{spie}  %>>> use this instead for A4 paper
\usepackage[]{graphicx}
\usepackage{amssymb,amsmath,psfrag,url}
\usepackage{verbatim}

\newcommand{\Bolivarallee}{Boliva\hspace{-0.1mm}r\hspace{0.15mm}a\hspace{-0.1mm}llee}

\newcommand{\Takustrasse}{Taku\hspace{0.25mm}s\hspace{-0.1mm}tra{\ss}e}

\newcommand{\Field}[1]{{\bf{#1}}}
\newcommand{\Tensor}[1]{{\bf{#1}}}

\newcommand{\myset}[1]{\left\{#1\right\}}

\newcommand{\real}{\mathbb{R}}

\newcommand{\curl}{\mathbf{curl}\;}
\newcommand{\curlz}{\mathbf{curl}}

\newcommand{\hcurl}{\mathrm{H}\left(\curlz\right)}

\title{hp-finite element method for simulating light scattering from complex 3D structures}

\author{
Sven~Burger,\supit{\,ab}
Lin~Zschiedrich,\supit{\,a}
Jan~Pomplun,\supit{\,a}
Sven~Herrmann,\supit{\,b}
Frank~Schmidt\supit{\,ab}
\skiplinehalf
\supit{a}
JCMwave GmbH,
\Bolivarallee~22, 
D\,--\,14\,050 Berlin,
Germany
\smallskip\\
\supit{b}
Zuse Institute Berlin\,(ZIB),
\Takustrasse~7,
D\,--\,14\,195 Berlin,
Germany
\authorinfo{
Corresponding author: S.~Burger\\
URL: http://www.jcmwave.com\\
URL: http://www.zib.de
}}

\begin{document}
\maketitle
%%%%%%%%%%%%%%%%%%%%%%%%%%%%%%%%%%%%%%%%%%%%%%%%%%%%%%%%%%%%% 
%% SPIE Copyright form 
\noindent
This paper will be published in Proc.~SPIE Vol.~{\bf 9424}
(2015) 94240Z, ({\it Metrology, Inspection, and Process Control for Microlithography XXIX}, DOI: 10.1117/12.2085795), 
and is made available 
as an electronic preprint with permission of SPIE. 
One print or electronic copy may be made for personal use only. 
Systematic or multiple reproduction, distribution to multiple 
locations via electronic or other means, duplication of any 
material in this paper for a fee or for commercial purposes, 
or modification of the content of the paper are prohibited.
Please see original paper for images at higher resolution. 
%%%%%%%%%%%%%%%%%%%%%%%%%%%%%%%%%%%%%%%%%%%%%%%%%%%%%%%%%%%%% 

\begin{abstract}
Methods for solving Maxwell's equations are integral 
part of optical metrology and computational lithography setups. 
Applications require accurate geometrical resolution, 
high numerical accuracy and/or low computation times. 
We present a finite-element based electromagnetic field solver relying on unstructured 
3D meshes and adaptive hp-refinement. 
We apply the method for simulating light scattering off arrays of high aspect-ratio 
nano-posts and FinFETs.
\end{abstract}


\section{Introduction}

\begin{figure}[b]
\begin{center}
\psfrag{r1}{\sffamily $r_\textrm{F}$}
\psfrag{r2}{\sffamily $r_\textrm{G}$}
\psfrag{Rs, Rp}{\sffamily R$_\textrm{S}$, R$_\textrm{P}$}
\psfrag{Wavelength [nm]}{\sffamily Wavelength\,[nm]}
\includegraphics[width=.6\textwidth]{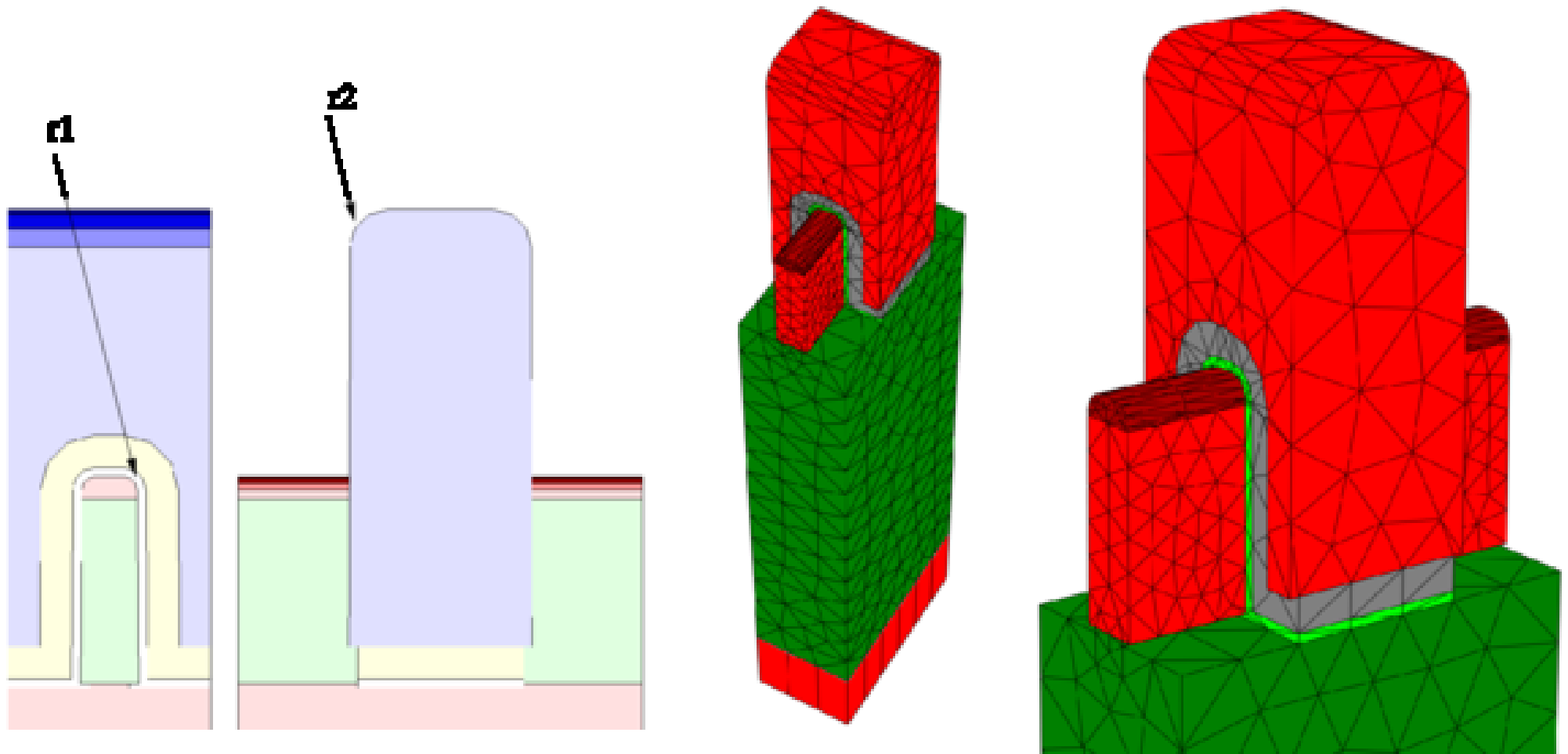}
\includegraphics[width=.35\textwidth]{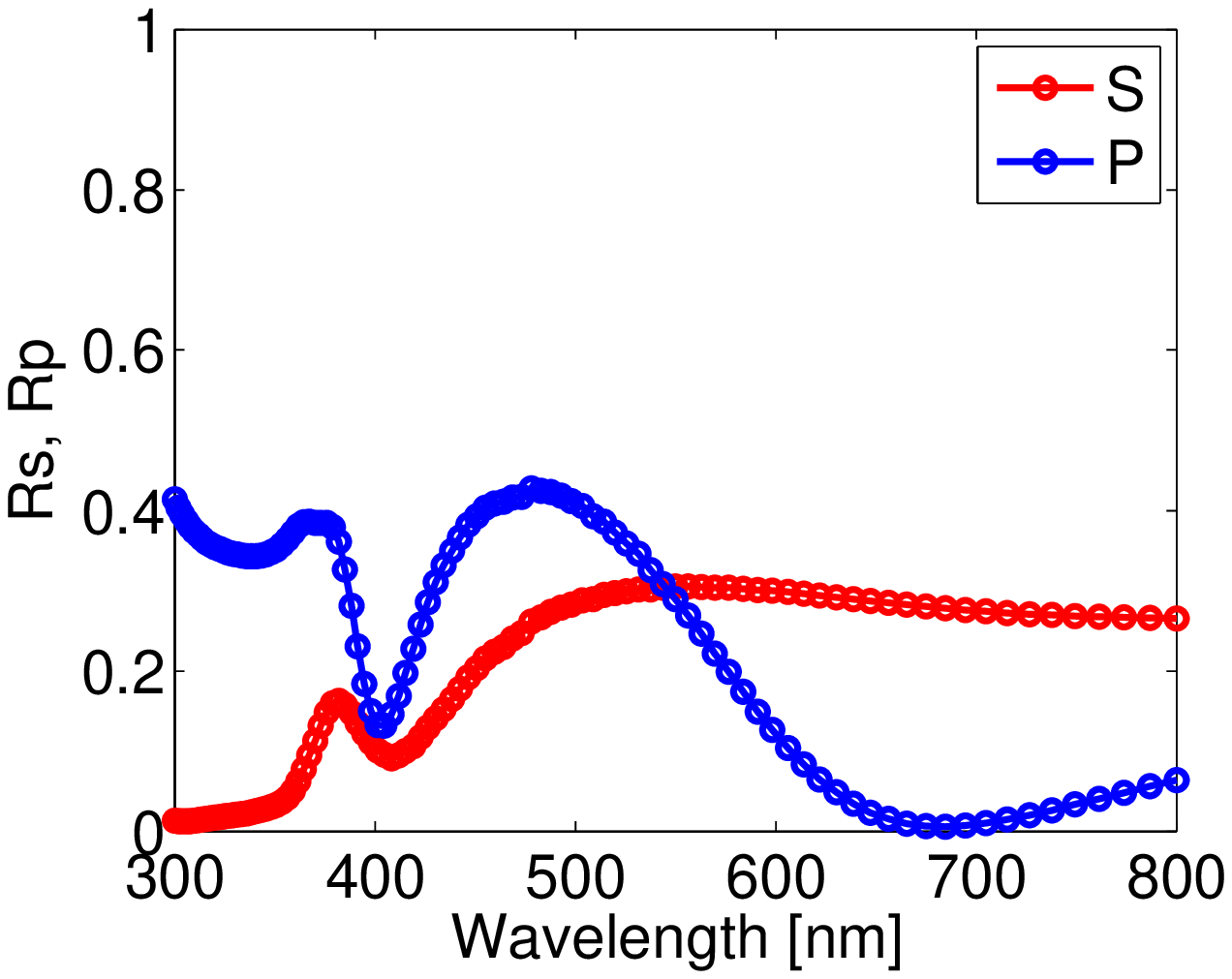}
  \caption{
FinFET: {\it Left:} Images of the geometry layout in two cross-sections. 
{\it Center:} Images of parts of the tetrahedral mesh in different viewing directions. 
{\it Right:} Computed reflection spectra for S- and P-polarized light. 
}
\label{fig_finfet_schematics}
\end{center}
\end{figure}
 
Optical metrology can be used to detect features sizes on a sub-nanometer level. 
In the semiconductor industry it is used in process control and in mask quality control 
for pushing performance of DUV and EUV lithography~\cite{Pang2012aot}. %,Lai2012aot}. 
Numerical modelling is an important part of optical metrology setups in this field: measurement results are 
compared to simulation results of a parameterized model in order to quantitatively determine dimensions of 
the measured sample.  
With increasing complexity and decreasing feature sizes, 
the need for accurate optical metrology methods for complex 3D shapes is increasing~\cite{Topol2006,Bunday2013spie}.
This triggers also a need for efficient numerical methods for computational metrology. 

The challenge for electromagnetic field (EMF) solvers (Maxwell solvers) 
is typically efficiency (i.e., to achieve highly accurate results at low computation times). 
Finite-element methods (FEM) allow for high efficiency due to accurate geometrical modelling, 
adaptive meshing strategies, and higher-order convergence. 
In simulation tasks requiring high accuracy FEM can outperform other rigorous simulation 
methods~\cite{Burger2005bacus,Burger2008bacus,Hoffmann2009spie,Maes2013oe}. 

We develop and investigate finite-element methods for electromagnetic field simulations. 
In previous contributions these have been applied to various setups in optical 
metrology~\cite{Scholze2007a,quintanilha2009critical,zang2011structural,Bodermann2012op,burger2013al,Soltwisch2013eom,Soltwisch2014pm,Petrik2015jeos}.
In this context, also 3D structures have been investigated~\cite{Burger2011eom1_,Burger2011pm1,Kleemann2011eom3,Kato2012a}.
Here we discuss methods for further performance improvements, especially for efficient 
simulation of 3D devices with complex geometries. 
This is reached by using hp-finite elements on unstructured, tetrahedral and prismatoidal meshes. 
Figure~\ref{fig_finfet_schematics} shows the model, mesh, and simulation result of a typical investigated sample.

This paper is structured as follows: 
The background of the numerical method is presented in Section~\ref{section_background}. 
The method is validated by presenting simulation results for two different examples  
related to current critical dimension (CD) metrology requirements~\cite{Bunday2013spie,Bodermann2012op}:
Section~\ref{section_nanoposts} presents simulations of nano-post arrays with high aspect-ratio. 
Section~\ref{section_finfet} presents simulations of arrays of fin field-effect transistors (FinFET).

\section{hp Finite element method}
\label{section_background}
In the following the background of the finite element method is summarized~\cite{MON03}.
Light scattering off nanoscopic structures on scatterometry samples is modeled by 
the linear Maxwell's equations in frequency domain~\cite{Pomplun2007pssb,Burger2012springer}. 
From these a single equation for the electric field $\Field{E}$ can be derived:
\begin{equation}
  \curl\Tensor{\mu}^{-1}\curl \Field{E}-\omega^{2}\Tensor{\epsilon}\Field{E}=i\omega\Field{J}.
  \label{eq:mwE}
\end{equation}
where  $\Tensor{\epsilon}$ and $\Tensor{\mu}$ are the permittivity and permeability tensor, $\omega$ is 
the time-harmonic frequency of the electromagnetic field, and the  
electric current $\Field{J}$ is source of an electromagnetic field. 
The domain of interest is separated into an infinite 
exterior $\Omega_{\mathrm{ext}}$ which hosts the given incident field and the scattered field, 
and an interior $\Omega_{\mathrm{int}}$ where the total field is computed. 
Electromagnetic waves incident from the exterior to the interior at the boundaries between both domains
are added to the right hand side of Eq.~\eqref{eq:mwE}. 
For numerical simulations the infinite exterior is treated using transparent 
boundary conditions (using the perfectly matched layer method, PML).

\begin{figure}[t]
\begin{center}
\psfrag{cd1}{\small \sffamily CD$_1$}
\psfrag{cd2}{\small \sffamily CD$_2$}
\psfrag{cd3}{\small \sffamily CD$_3$}
\psfrag{h1}{\sffamily $h_1$}
\psfrag{h2}{\sffamily $h_2$}
\psfrag{px}{\sffamily $p_x$}
\psfrag{py}{\sffamily $p_y$}
\psfrag{Rs}{\sffamily R$_\textrm{S},p$}
\psfrag{Rp}{\sffamily R$_\textrm{P},p$}
\psfrag{Wavelength [nm]}{\sffamily Wavelength\,[nm]}
  \includegraphics[width=.15\textwidth]{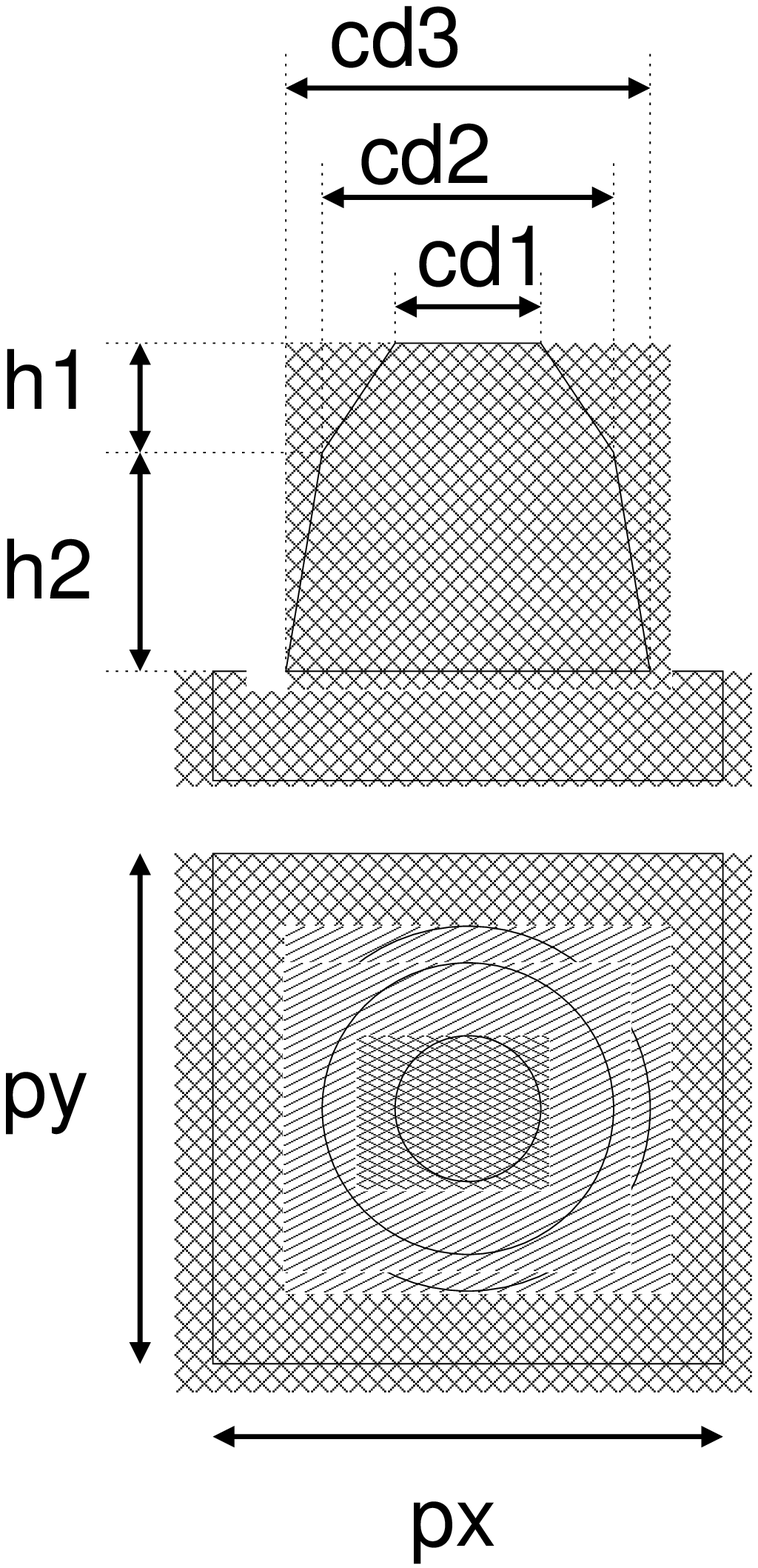}
  \hspace{2mm}
  \includegraphics[width=.39\textwidth]{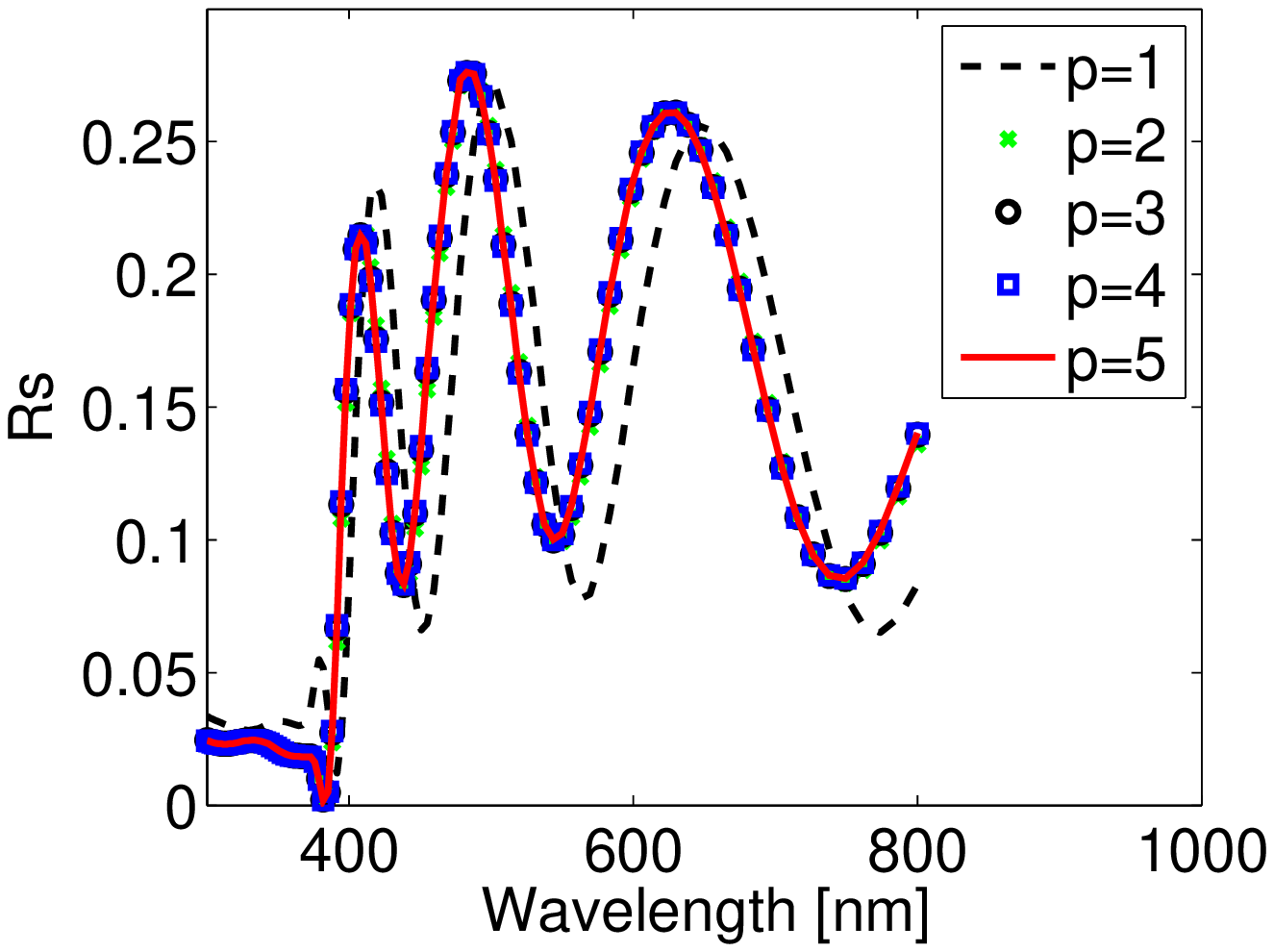}
  \includegraphics[width=.39\textwidth]{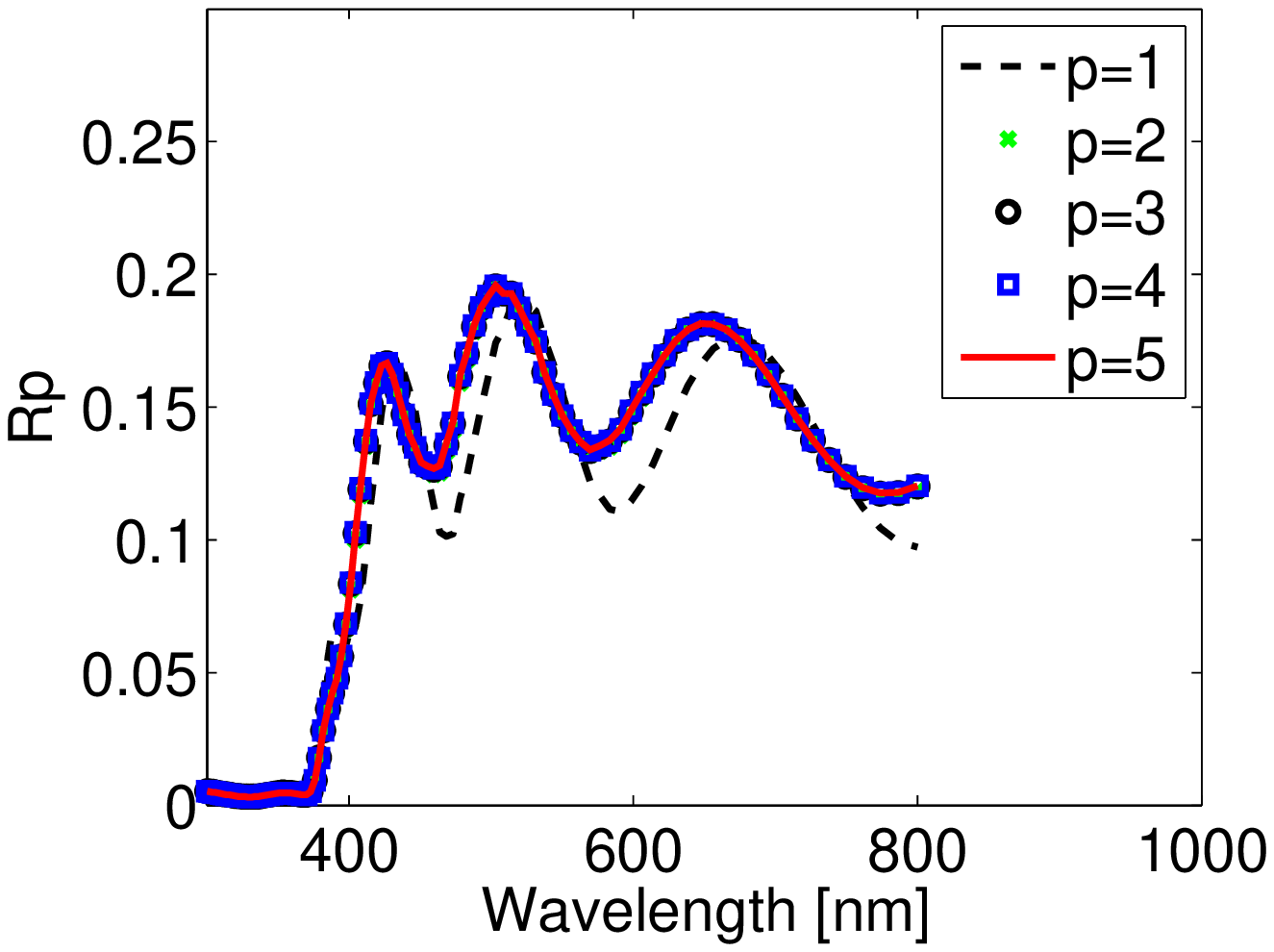}
  \caption{
{\it Left:} 
Schematics of the geometry of the investigated scatterometric target (unit cell of a 2D-periodic grating of nano-posts with 
circular cross section, {\it not to scale}). 
Different parameters of the model are indicated (critical dimension, CD, at different heights, $h_1$, $h_2$, pitches $p_x$, $p_y$). 
{\it Center / right:} Reflection spectra of S/P-polarized light, $R_{\textrm{S/P}, p}$,  for numerical discretization parameters 
$p=1...5$.
}
\label{schematics_nanopost}
\label{fig_nanopost_rsp}
\end{center}
\end{figure}

For a FEM discretization,  Eq.~\eqref{eq:mwE} is first transformed into a weak formulation, i.e., 
it is tested with a vectorial function $\phi$ and integrated over $\real^{3}$ which yields:

\begin{equation}
\int\limits_{\real^{3}}\curl \phi \;\Tensor{\mu}^{-1}\curl \Field{E}-\omega^{2}\phi\,\Tensor{\varepsilon}\Field{E}=i\omega\int\limits_{\real^{3}}\phi\,\Field{J}.
   \label{eq:1}
\end{equation}

For compact notation, the forms $a(\phi,\Field{E})$ and $f(\phi)$ are introduced, 
and 
the function space $\hcurl$ is defined \cite{MON03}.
The weak form of Maxwell's equations then reads:
\\\noindent
Find $\Field{E}\in\hcurl$ such that:
\begin{equation}
  \label{eq:mwEweak}
a(\phi,\Field{E})=f(\phi)\,,\quad\forall \phi\in\hcurl.
\end{equation}

A finite element discretization of Maxwell's equations restricts the 
formulation \eqref{eq:mwEweak} to a finite-dimensional subspace $V_h$ with $\dim V_h=N<\infty$:\\\noindent
Find $\Field{E}_h\in V_h$ such that:
\begin{equation}
  \label{eq:mwEweakFEM}
a(\phi_h,\Field{E}_h)=f(\phi_h)\,,\quad\forall \phi_h\in V_h.
\end{equation}

Next, a basis $\myset{\varphi_1,\dots,\varphi_N}$ 
of $V_h$ is constructed, and the electric field is expanded using the basis elements: $\Field{E}_h=\sum\limits_{i=1}^N e_i \varphi_i$. 
The variational problem (Eq.~\ref{eq:mwEweakFEM}) is then tested with all elements of the basis which gives a linear system of equations:
\begin{equation}
\sum\limits_{i=1}^N a(\varphi_j,\varphi_i)e_i=f(\varphi_j)\,,\quad\forall j=1,\dots,N.
\label{equation_matrix}
\end{equation}
The matrix $A_{ji}=a(\varphi_j,\varphi_i)$ is sparse and can be decomposed with efficient sparse 
LU solvers to obtain the unknown expansion coefficients $e_i$ of the electric field. 

\begin{figure}[t]
\begin{center}
\psfrag{Err(Rs)}{\sffamily $\Delta\textrm{R}_{\textrm{S},p}$}
\psfrag{Err(Rp)}{\sffamily $\Delta\textrm{R}_{\textrm{P},p}$}
\psfrag{Wavelength [nm]}{\sffamily Wavelength\,[nm]}
\includegraphics[width=.4\textwidth]{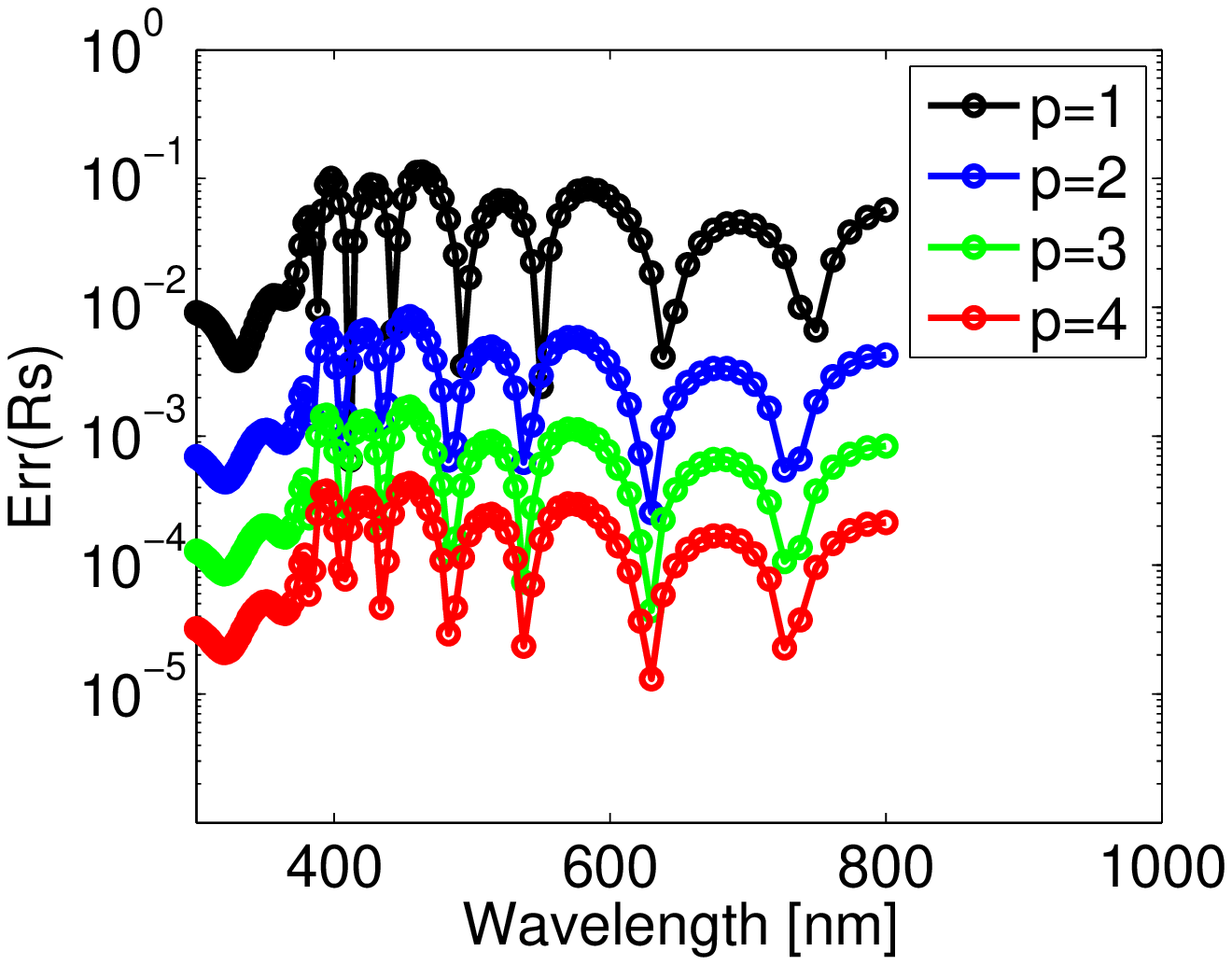}
\includegraphics[width=.4\textwidth]{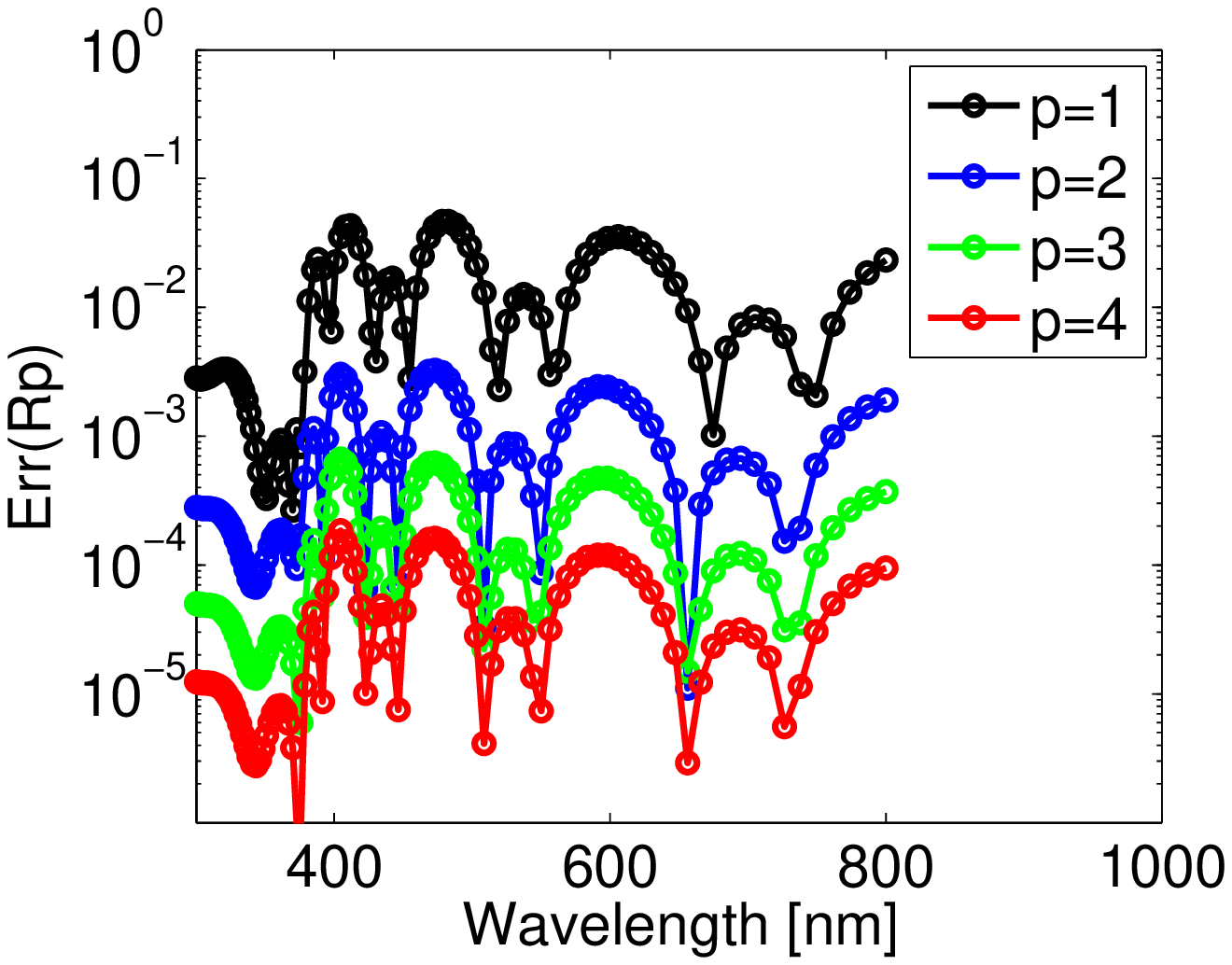}
  \caption{
Spectral dependence of numerical discretization error $\Delta\textrm{R}_{\textrm{S/P},p}$ for S/P polarization, obtained for
various horizontal FEM degrees $p$, with fixed $p_z=5$.
}
\label{fig_nanopost_err_rsp}
\end{center}
\end{figure}

The basis $\myset{\varphi_1,\dots,\varphi_N}$  is constructed using elements $\varphi_i$ (also called ansatz functions) 
which are polynomial functions of 
order $p$, and which are defined on a single patch of the spatial discretization of the geometry (mesh) only. 
For the results presented in this paper, we attribute elements of different polynomial order $p$ to different patches of the 
mesh~\cite{Babuska1981}. 
In regions where the mesh is very fine due to required geometry resolution (very thin layers or other fine details of the geometry)
a lower polynomial order $p$ can be chosen than in regions where the mesh is coarser. 
The method to distribute different orders $p$ to the different patches relies on estimating errors on the different patches, making use of  
informations on geometry, meshing, material properties and source fields. 
This yields a basis $\myset{\varphi_1,\dots,\varphi_N}$ which is well adapted to the problem and does not spend too much computational 
effort in regions where it is not required. 

We demonstrate this so called {\it hp}-FEM method for two different applications: 
an array of very elongated posts and a transistor geometry with nanometer features (FinFET). 
For the array of nano-posts, due to the elongated geometry, best performance is reached when the mesh consists of 
elongated elements (in this case prismatoidal elements) and when different polynomial degrees $p$ are used in the dimension of elongation and 
in the orthogonal dimensions. 
For the transistor, an unstructured tetrahedral mesh discretizes the geometry, where the typical dimensions 
of the mesh elements, $h$, can vary over about two orders of magnitude. In this case different polynomial degrees $p$ are used 
for different tetrahedral elements. 

%%%%%%%%%%%%%%%%%%%%%%%%%%%%%%%%%%%%%%%%%%%%%%%%%%%%%%%%%%%%%%%%%%%%%%%%%%%%%%
% NANO POST
%%%%%%%%%%%%%%%%%%%%%%%%%%%%%%%%%%%%%%%%%%%%%%%%%%%%%%%%%%%%%%%%%%%%%%%%%%%%%%

%%%%%%%%%%%%%%%%%%%%%%%%%%%%%%%%%%%%%%%%%%%%%%%%%%%%%%%%%%%%%%%%%%%%%%%%%%%%%%
\section{Simulation of light scattering off silicon nano-posts with high aspect-ratio}
\label{section_nanoposts}
The model investigated in this section corresponds to a periodic array of Silicon nano-post on a silicon substrate. 
The geometrical setup is described schematically in Fig.~\ref{schematics_nanopost} (left), and the 
parameter configuration for the simulations is defined in Table~\ref{table_specs_nanopost}.

\begin{table}[h]
\begin{center}
\begin{tabular}{|l|l|}
\hline
%material & Si \\ \hline
$p_x=p_y$ & 32\,nm\\ \hline
CD$_1$ / CD$_2$ / CD$_3$ & 16\,nm / 22\,nm / 24\,nm\\\hline
$h_1$ /$h_2$ & 300\,nm / 400\,nm\\ \hline
%$\lambda$& 300\,nm -- 800\,nm \\ \hline
$\theta$ / $\phi$ & 30\,deg /  0\,deg \\ \hline
\end{tabular}
\caption{Parameter settings for the Si nano-post array (compare Fig.~\ref{schematics_nanopost}).
}
\label{table_specs_nanopost}
\end{center}
\end{table}

\begin{figure}[b]
\begin{center}
\psfrag{Average error}{\sffamily $\overline{\Delta\textrm{R}_{p}}$}
\psfrag{Av. error}{\sffamily $\overline{\Delta\textrm{R}_{N_c}}$}
\psfrag{p}{\sffamily $p$}
\psfrag{Nc}{\sffamily $N_c$}
\includegraphics[width=.4\textwidth]{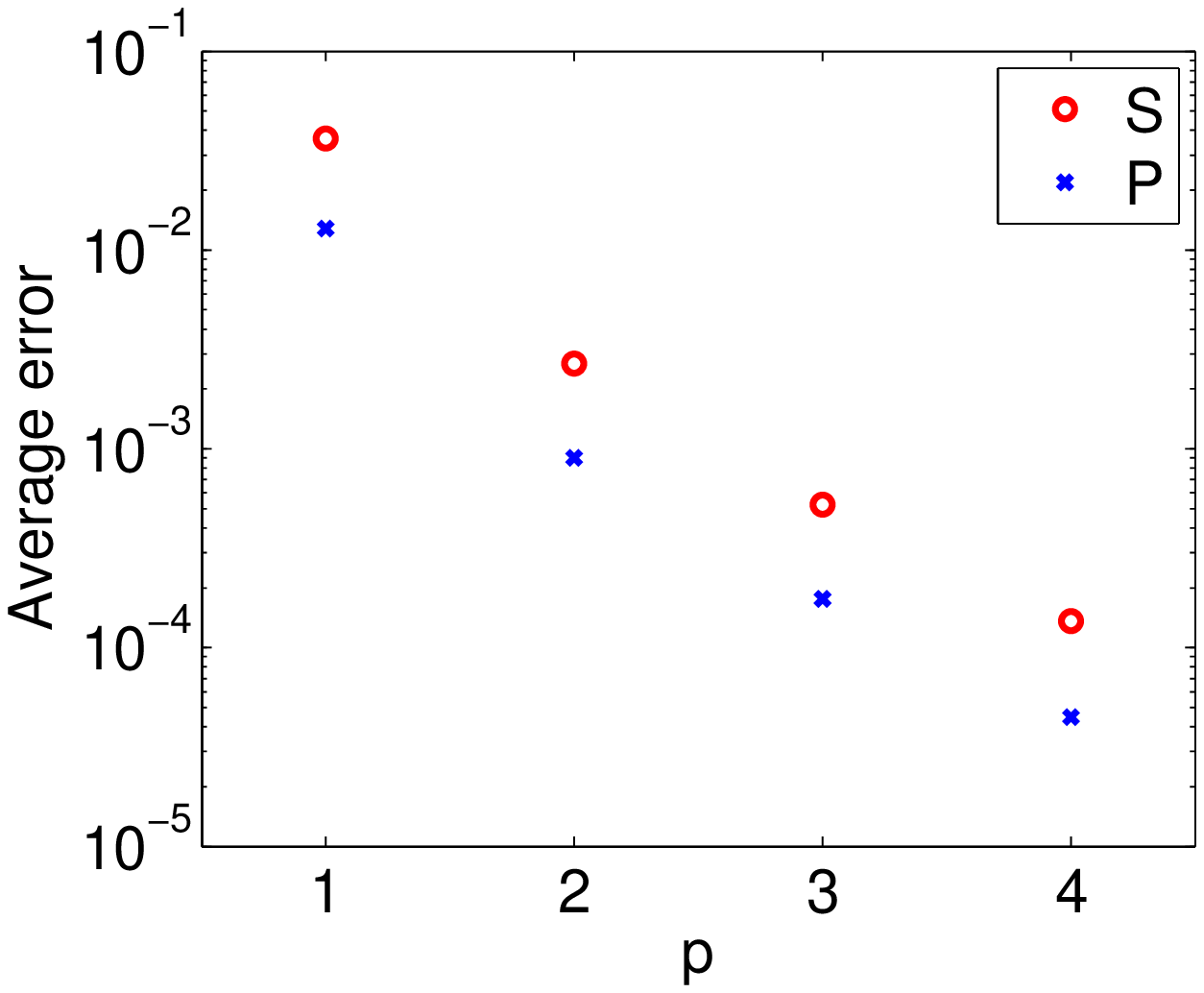}
\includegraphics[width=.4\textwidth]{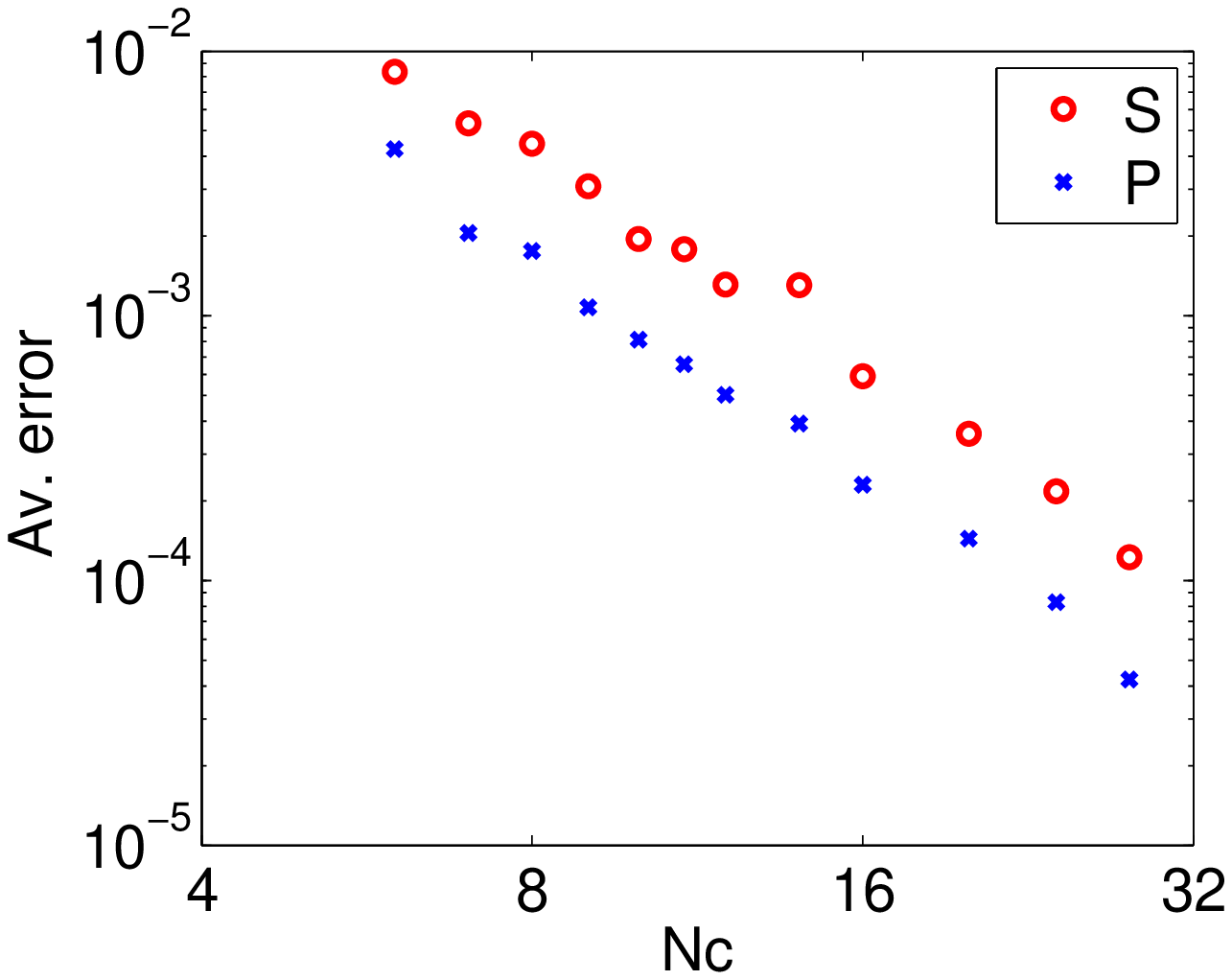}
  \caption{
{\it Left:} Convergence of the average error $\overline{\Delta\textrm{R}_{p}}$ with horizontal finite element degree $p$, 
for S- and P-polarized incident light.
{\it Right:} Convergence of the average error $\overline{\Delta\textrm{R}_{p}}$ with (geometrical) discretization parameter $N_c$
(number of segments of the polygon defining the cross-section of the nano-post).
}
\label{fig_nanopost_conv}
\end{center}
\end{figure}

In the model the structure is illuminated from above (superspace with refractive index $n=1$) 
with S- and P-polarized plane waves, at an angle $\theta$
to the surface normal, and rotation angle $\phi$.
For simulating a wavelength spectrum, independent computations of the time-harmonic model for 100 wavelengths 
are performed. 
The optical material parameters of Silicon at each wavelength are obtained from tabulated data~\cite{Palik1985}.
In our setup, up to 80 independent computations are processed in parallel on the computation threads of a standard workstation. 
Typical computation times for a single computation at low to medium accuracy levels range from below 1\,sec to few seconds. 

Fig.~\ref{fig_nanopost_rsp} (center, right) shows spectra of reflected S- and P-polarized light, R$_\textrm{S}$ and R$_\textrm{P}$. 
The displayed spectra are obtained for the same physical setting, however, at different numerical resolutions $p$, i.e., R$_{\textrm{S},p}$. 
In this case, we always use the same spatial mesh where the geometry is discretized using prismatoidal elements which are 
elongated in $z$-direction (surface normal). 
The numerical discretization parameter $p$ here corresponds to the polynomial order of the finite-element ansatz functions 
$\varphi_i$ (cf., Sec.~\ref{section_background}) in the $x-y$-plane. The polynomial order of the ansatz functions 
in $z$-direction in this case is chosen as $p_z = 5$.
For the absolute values on a linear scale, as displayed in Fig.~\ref{fig_nanopost_rsp}, differences in the computed spectra 
for $p>2$ can hardly be detected. 
Therefore, in Figure~\ref{fig_nanopost_err_rsp} the differences of the computed spectra to the spectra computed at 
highest numerical resolution, $\Delta\textrm{R}_{\textrm{S},p} = |\textrm{R}_{\textrm{S},p} - \textrm{R}_{\textrm{S},p=5}|$, 
are displayed on a logarithmic scale. 
$\Delta\textrm{R}$ is also termed numerical discretization error.
As can be expected the numerical discretization error decreases exponentially with $p$ in the whole spectral range. 
We define the average discretization error as $\overline{\Delta\textrm{R}_{p}}=\sum_1^N\Delta\textrm{R}_{p}/N$,
where the summation is performed over all $N$ spectral points. 
Please note that several different conventions are used for defining numerical discretization errors in this 
context. 
The average error obtained from the data in Fig.~\ref{fig_nanopost_err_rsp} is displayed in 
Fig.~\ref{fig_nanopost_conv} (left). 
In the investigated parameter regime, exponential convergence to very high accuracy levels is observed. 

All results displayed in Figures~\ref{fig_nanopost_rsp} to~\ref{fig_nanopost_conv} (left)
are obtained on the same discretization of the geometry (essentially on the same prismatoidal mesh). 
In order to verify accuracy of the discrete geometry we have performed a convergence study for a meshing parameter:
For meshing the nano-post geometry the mesh generator automatically discretizes the circular cross-section of 
the nano-post with a polygon, with a given number of segments, $N_c$. 
We have computed reflectivity spectra similar as in Fig.~\ref{fig_nanopost_rsp} for different segment numbers $N_c$. 
Here we have defined the average discretization error as 
$\overline{\Delta\textrm{R}_{N_c}}=\sum_1^N\Delta\textrm{R}_{N_c}/N$
with 
$\Delta\textrm{R}_{N_c} = |\textrm{R}_{N_c} - \textrm{R}_{N_c=36}|$.
As can be seen from Fig.~\ref{fig_nanopost_conv} (right) the results converge very well with geometry discretization 
parameter $N_c$. Results with an accuracy $\overline{\Delta\textrm{R}_{N_c}}<0.2\%$ are obtained for $N_c\ge 10$.

\begin{figure}[b]
\begin{center}
\psfrag{Pre}{\sffamily $p_\textrm{Prec}$}
\psfrag{Fraction of elements}{\sffamily Fraction of elements}
\psfrag{Average error}{\sffamily $\overline{\Delta\textrm{R}_{p_\textrm{Prec}}}$}
\includegraphics[width=.4\textwidth]{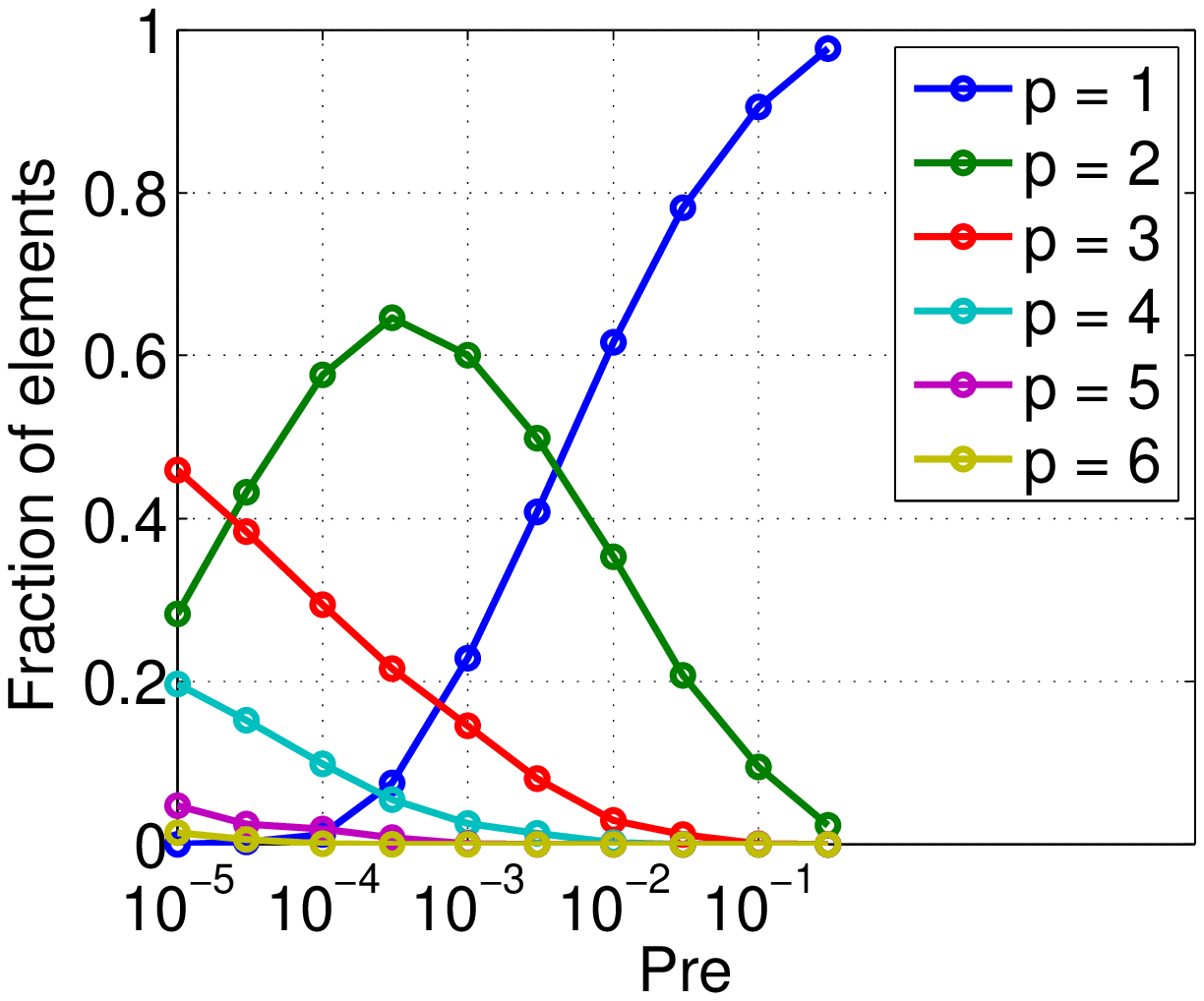}
\hspace{5mm}
\includegraphics[width=.4\textwidth]{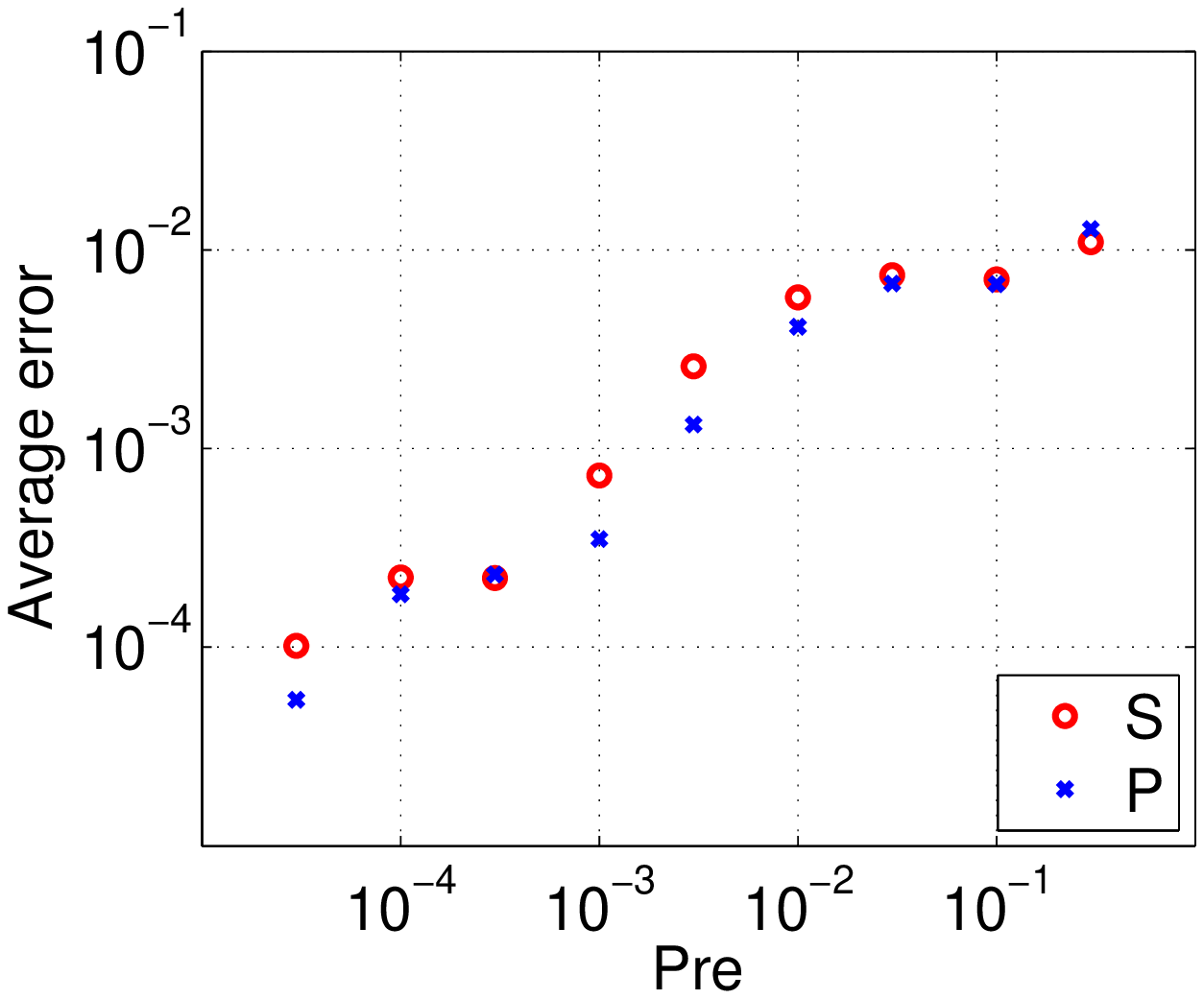}
  \caption{
Hp-adaptive FinFET simulations: 
{\it Left:} 
Fraction of mesh elements on which ansatz functions $\varphi_i$ with polyomial order $p$ are used, as a function of 
the numerical accuracy parameter $p_\textrm{Prec}$. 
{\it Right:} Convergence of average numerical error $\overline{\Delta\textrm{R}_{p_\textrm{Prec}}}$ with accuracy parameter $p_\textrm{Prec}$.
}
\label{fig_finfet_conv}
\end{center}
\end{figure}

%%%%%%%%%%%%%%%%%%%%%%%%%%%%%%%%%%%%%%%%%%%%%%%%%%%%%%%%%%%%%%%%%%%%%%%%%%%%%%
% FINFET
%%%%%%%%%%%%%%%%%%%%%%%%%%%%%%%%%%%%%%%%%%%%%%%%%%%%%%%%%%%%%%%%%%%%%%%%%%%%%%

%%%%%%%%%%%%%%%%%%%%%%%%%%%%%%%%%%%%%%%%%%%%%%%%%%%%%%%%%%%%%%%%%%%%%%%%%%%%%%

\section{Simulation of light scattering off FinFETs}
\label{section_finfet}
The model investigated in this section corresponds to a periodic array of FinFET structures of the 22\,nm (and smaller) 
technology nodes.
Bunday {\it et al}~\cite{Bunday2013spie} point out that with the launch of such small structures on integrated circuits 
complex 3D architectures have become a cruicial driver for down-scaling, and that this 
also implies additional needs for metrology.  

\begin{figure}[t]
\begin{center}
\includegraphics[width=.7\textwidth]{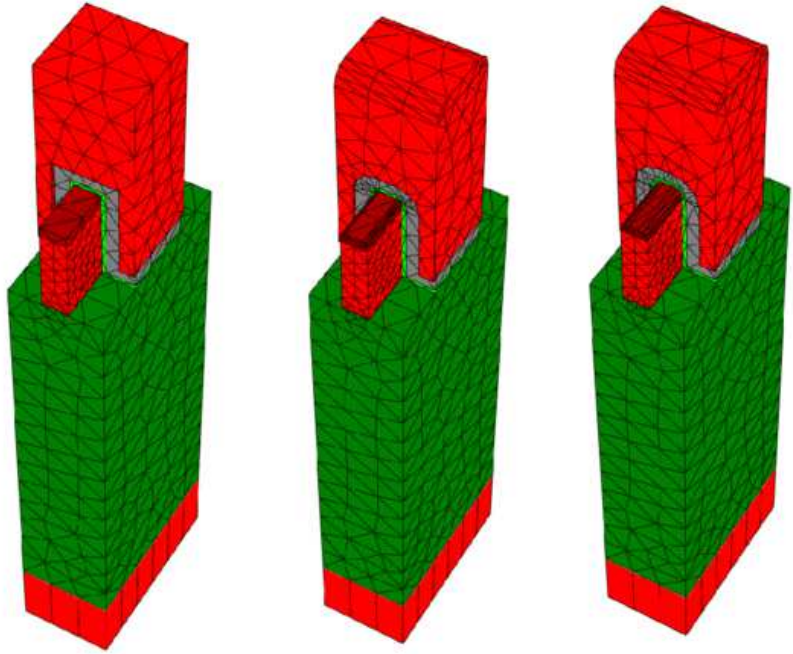}
  \caption{
Tetrahedral meshes of the FinFET geometry with different settings of corner roundings:
{\it Left:} No corner rounding, $r_\textrm{F}=0$, $r_\textrm{G}=0$.
{\it Center:} $r_\textrm{F}=1\,nm$, $r_\textrm{G}=8\,nm$.
{\it Right:} $r_\textrm{F}=4\,nm$, $r_\textrm{G}=8\,nm$.
}
\label{fig_finfet_r_meshes}
\end{center}
\end{figure}

\begin{figure}[b]
\begin{center}
\psfrag{r=0}{\sffamily \small $r:0/0$nm}
\psfrag{r=3/8nm}{\sffamily \footnotesize $r:3/8$nm}
\psfrag{dRp}{\sffamily $dR_\textrm{P}$}
\psfrag{Wavelength [nm]}{\sffamily Wavelength\,[nm]}
\includegraphics[width=.4\textwidth]{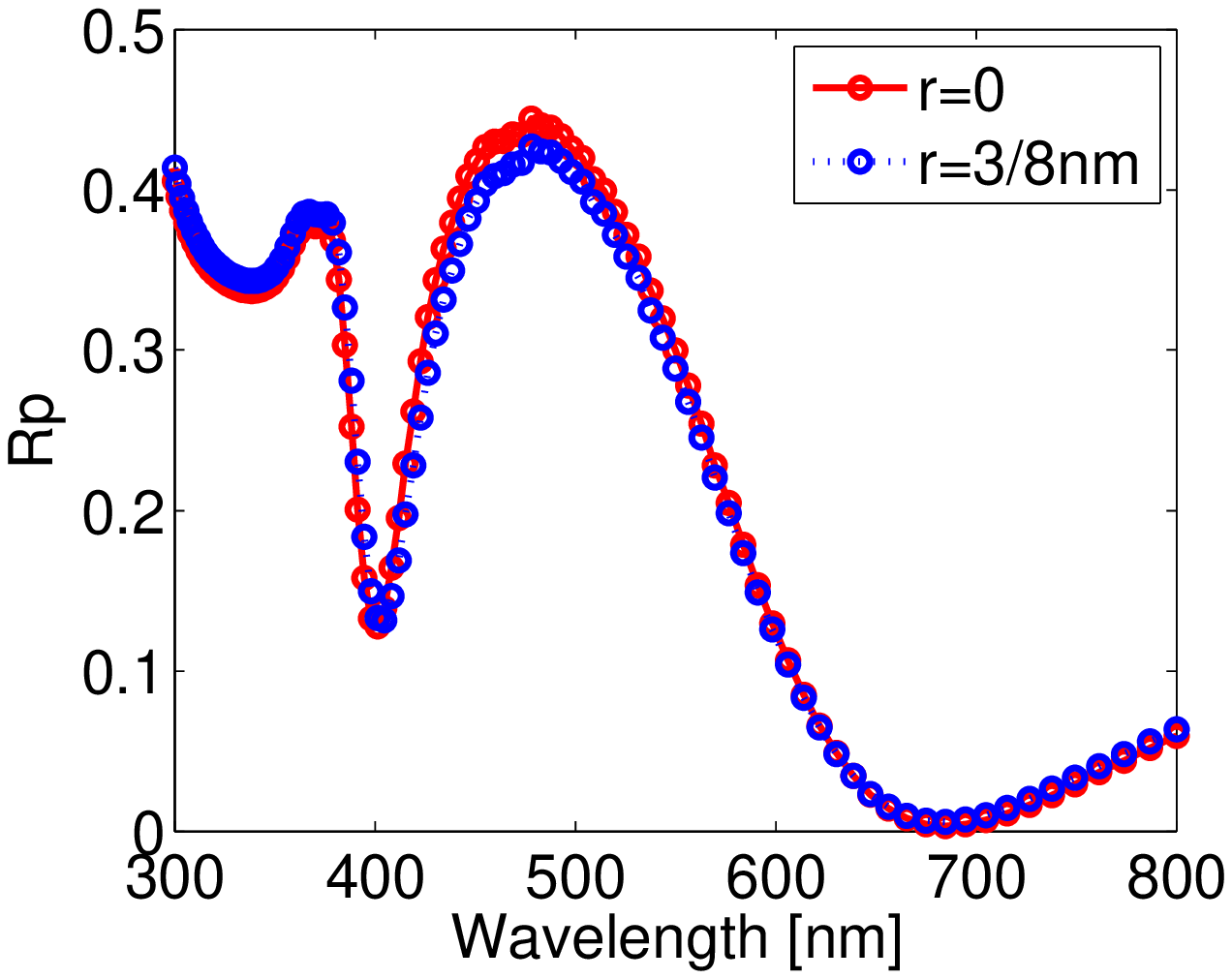}
\includegraphics[width=.4\textwidth]{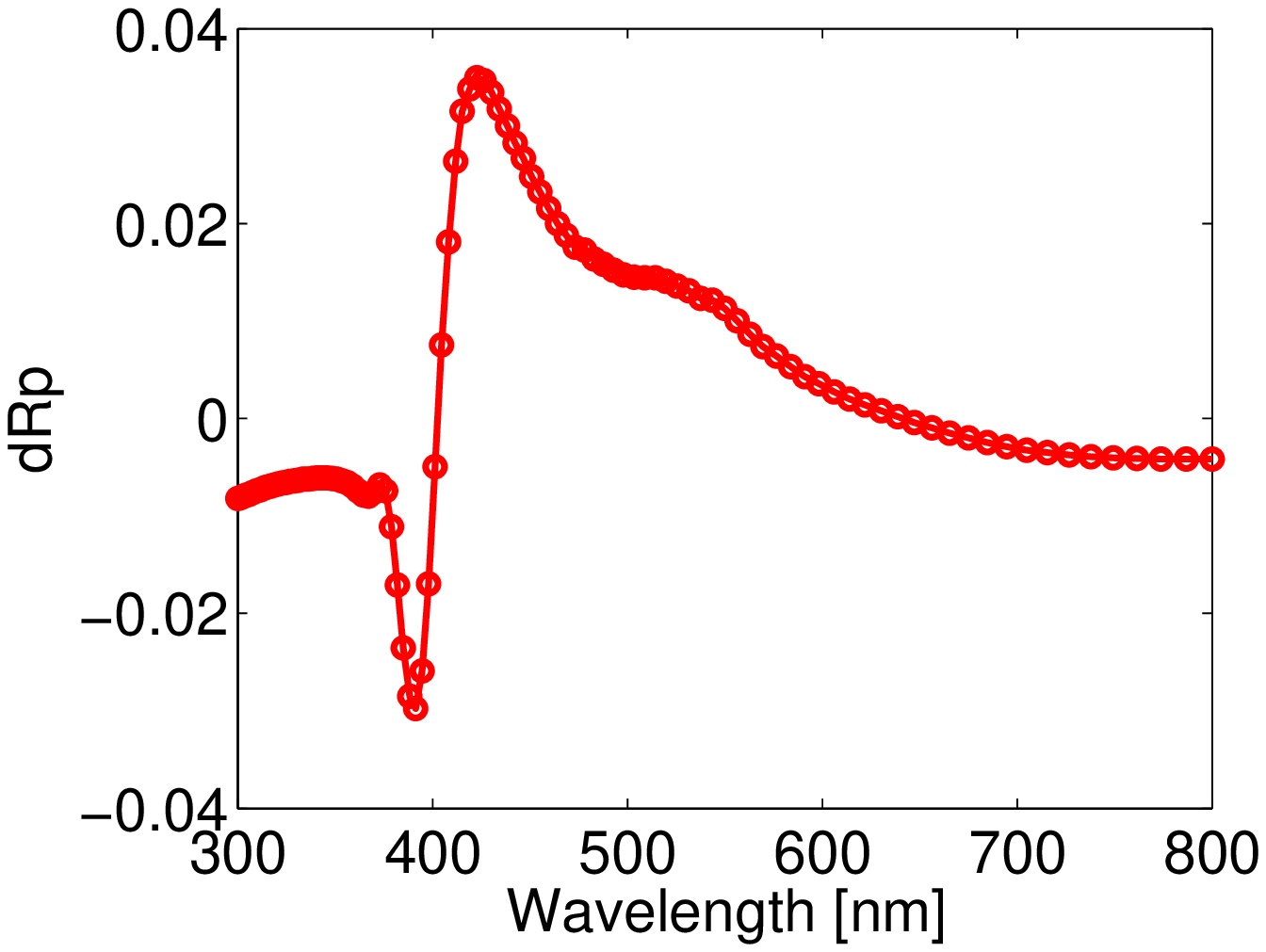}
%\hspace{5mm}
  \caption{
{\it Left:} 
Reflection spectra for P-polarized light, without corner rounding ($r_\textrm{F}=0$, $r_\textrm{G}=0$)
and with corner rounding ($r_\textrm{F}=3$\,nm, $r_\textrm{G}=8$\,nm).
{\it Right:} 
Difference between the two reflection spectra, 
$
dR_\textrm{P}=R_\textrm{P}(r_\textrm{F}=0, r_\textrm{G}=0) - R_\textrm{P}(r_\textrm{F}=3\textrm{nm}, r_\textrm{G}=8\textrm{nm})$.
}
\label{fig_finfet_r0_r3}
\end{center}
\end{figure}

The investigated geometrical setup is schematically shown in Fig.~\ref{fig_finfet_schematics} (left). 
All dimensions of the device follow Fig.~15 and Table~10 of Bunday {\it et al}~\cite{Bunday2013spie}
(22\,nm node: Fin pitch: 44\,nm, Gate pitch: 88\,nm, Fin width: 12.7\,nm, Fin SWA: 89.5\,deg,
Fin height: 40\,nm, Fin undercut: 2.1\,nm, Gate width: 40\,nm, Gate SWA: 89.8\,deg, 
Gate height: 95\,nm, Gate undercut: 2\,nm, SiN thickness: 5\,nm, High-k layer thickness: 2\,nm, 
TiN thickness: 7\,nm, BOX thickness: 200\,nm). Additional parameters are roundings of 
the fin ($r_\textrm{F}$: 3\,nm) and of the gate ($r_\textrm{G}$: 8\,nm) top edges.
The optical material parameters in the investigated spectral range are again obtained 
from tabulated data~\cite{Palik1985}
(Si, SiN, SiO$_2$, TiN, TiO$_2$).

Fig.~\ref{fig_finfet_schematics} (center) shows parts of tetrahedral meshes discretizing the FinFET geometry. 
Meshing is performed with an automatic mesh-generator~\cite{Blome2014compel}, which is a part of the 
finite-element package. 
The unstructured mesh allows to accurately resolve fine geometrical features like sub-nm undercuts, 
sidewall-angles and corner-roundings with only few additional mesh elements, 
facilitating accurate geometry modelling. 
When high refractive index contrasts between the different involved materials are present, such 
accurate geometry resolution is essential for precise approximation of the electro-magnetic 
field distributions. 
This  enables accurate light scattering results in the far field, 
as inaccurate near-field resolution is generally not {\it ''smoothed out''} in the far field. 
Computed reflection spectra for S- and P-polarized incident light at oblique angle of incidence are 
displayed in Fig.~\ref{fig_finfet_schematics} (right).

The shortest dimensions of the edges of the tetrahedra in the meshes displayed in Fig.~\ref{fig_finfet_schematics} (center) 
are comparable to the smallest geometrical features, i.e., typically smaller than 1\,nm. However, for best performance, 
in regions of larger 
geometrical structures (e.g., in the gate, or in the buried oxide layer, etc.) mesh element dimensions are rather 
scaled with some fractions of the wavelength of light, i.e., several orders of magnitude larger than the dimensions
of the smallest mesh elements. 

When asssembling the finite-element matrix  $A_{ji}=a(\varphi_j,\varphi_i)$ ({\it cf.} Eq.~\ref{equation_matrix}), 
a-priori error-estimation is used to choose the individual polynomial order $p$ of the set of ansatz functions $\varphi_i$
on each individual mesh element. A parameter $p_\textrm{Prec}$ controls error-estimation. 
Figure~\ref{fig_finfet_conv} (left) shows how with different settings of $p_\textrm{Prec}$ the fraction of 
mesh elements where different polynomial orders $p$ are used is changing. 
E.g., for a setting of $p_\textrm{Prec}=10^{-2}$ on about 60\% of the mesh elements, ansatz functions $\varphi_i$ with polyomial order 
$p=1$ are used, on about 35\% of the mesh elements, ansatz functions with $p=2$ are used, and on the remaining elements, 
ansatz functions with $p\ge 3$ are used. 
With decreasing $p_\textrm{Prec}$ these percentages shift towards higher $p$ for the individual patches.
Therefore, the dimension of $A_{ji}$ increases with decreasing $p_\textrm{Prec}$,
numerical discretization errors are expected to decrease with $p_\textrm{Prec}$, 
and computational costs increase with $p_\textrm{Prec}$, however, only moderately~\cite{MON03}. 
Figure~\ref{fig_finfet_conv} (right) shows convergence of the computed spectrum with $p_\textrm{Prec}$: 
very high accuracy levels are obtained.

\begin{figure}[t]
\begin{center}
\psfrag{dRs}{\sffamily $dR_\textrm{S}$}
\psfrag{dRp}{\sffamily $dR_\textrm{P}$}
\psfrag{Wavelength [nm]}{\sffamily Wavelength\,[nm]}
\includegraphics[width=.4\textwidth]{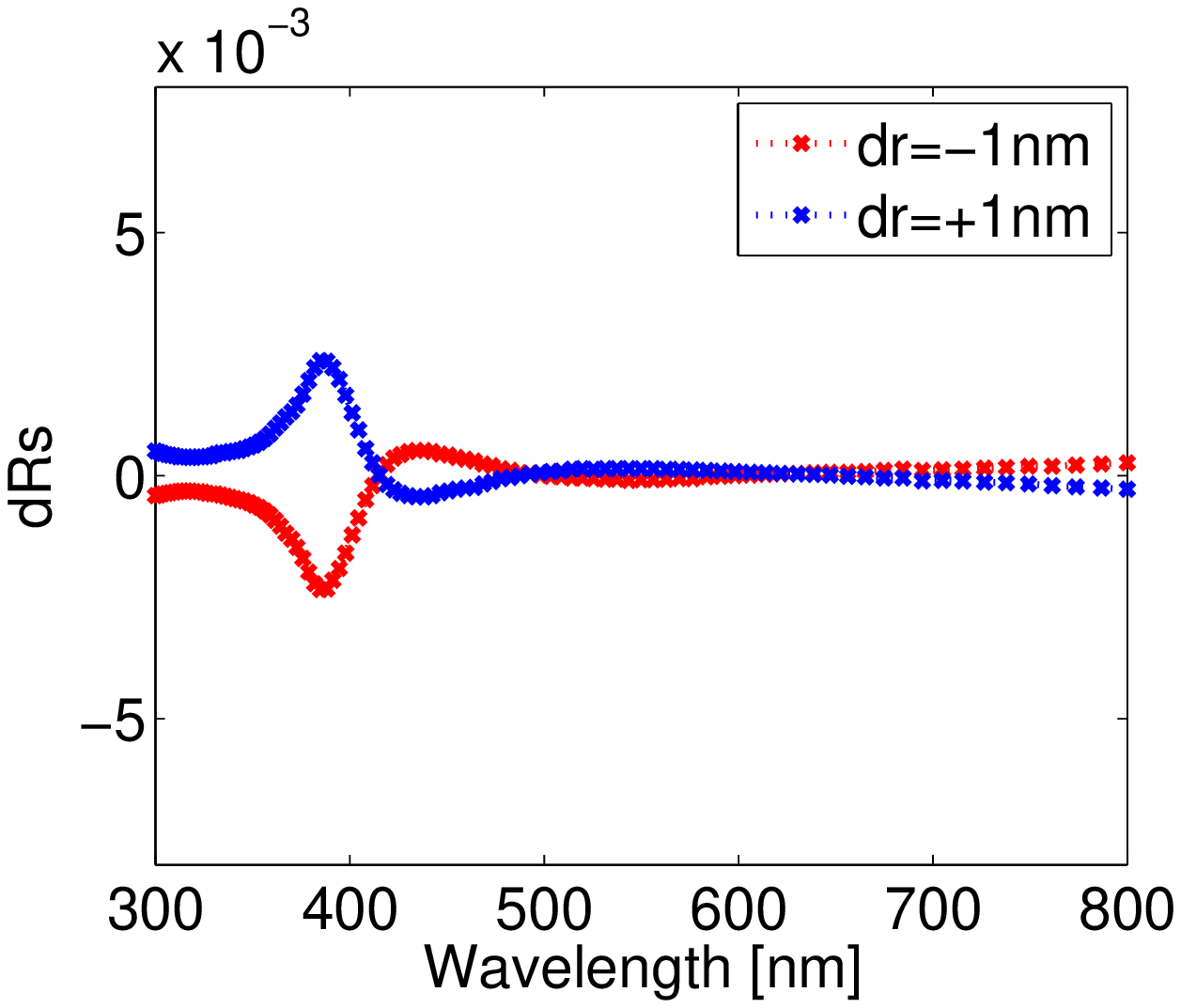}
\includegraphics[width=.4\textwidth]{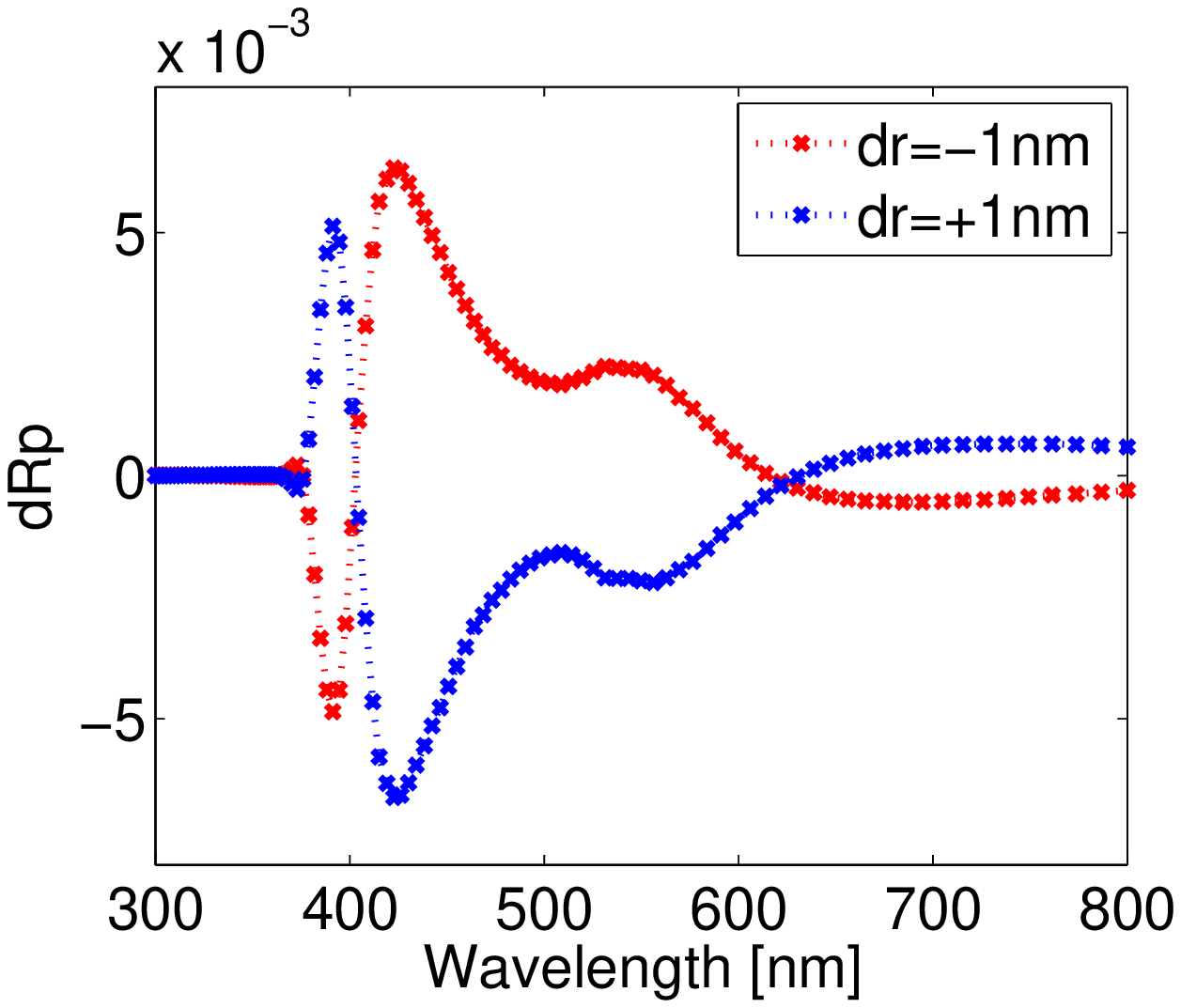}
%\hspace{5mm}
  \caption{
Spectral sensitivity of reflection of S ({\it left}) and P  ({\it right}) polarized 
light for changes of fin corner rounding radius $r_\textrm{F}$ around the value of $r_\textrm{F}=3$\,nm. 
}
\label{fig_finfet_dr}
\end{center}
\end{figure}

Unstructured meshes and finite elements with adaptive (hp) choice of polynomial order $p$ allows
to obtain fast and accurate results. 
This allows to analyze and reconstruct fine geometry features for complex, multi-scale geometries. 
To demonstrate this we investigate the influence of corner rounding on scattering spectra. 
Bunday {\it et al}~\cite{Bunday2013spie} note fin corner rounding as a parameter which does influence
FinFET performance. 
Figure~\ref{fig_finfet_r0_r3} compares spectra for FinFET geometries with and without fin and gate corner roundings. 
Figure~\ref{fig_finfet_dr} shows the changes of $S$ and $P$ reflection spectra with fin corner rounding radius $r_{\textrm F}$.  
In these simulations, gate rounding was fixed to $r_\textrm{G}=8$\,nm, and fin rounding was varied by 1\,nm around 
a rounding radius of $r_\textrm{F}=3$\,nm. 
The investigations can be used to design the scatterometric measurement setup for best sensitivity for 
specific parameters of interest~\cite{Soltwisch2013eom}. 
We note that sensitivities can also be computed directly at high accuracies~\cite{burger2013al}.
Values of $dR$ in Fig.~\ref{fig_finfet_dr} are of the order of up to few times $10^{-3}$ for  
changes of the rounding radius of 1\,nm.
This suggests that comparable or higher numerical accuarcy ranges should be used in reconstructions of 
such parameters.

%%%%%%%%%%%%%%%%%%%%%%%%%%%%%%%%%%%%%%%%%%%%%%%%%%%%%%%%%%%%%%%%%%%%%%%%%%%%%%
% CONCLUSION ETC
%%%%%%%%%%%%%%%%%%%%%%%%%%%%%%%%%%%%%%%%%%%%%%%%%%%%%%%%%%%%%%%%%%%%%%%%%%%%%%

%%%%%%%%%%%%%%%%%%%%%%%%%%%%%%%%%%%%%%%%%%%%%%%%%%%%%%%%%%%%%%%%%%%%%%%%%%%%%%

\section{Conclusion}
A finite-element method using hp-adaptivity on 3D prismatoidal and tetrahedral meshes has been demonstrated. 
Convergence to highly accurate results over a wide spectral range has been observed for examples related to
CD metrology challenges. 
The method allows to efficiently compute scatterometric signals and sensitivities. 

\section*{Acknowledgments}
The presented work  is part of the EMRP Joint Research Project IND\,17 {\sc Scatterometry}.
The EMRP is jointly funded by the EMRP participating countries within EURAMET and the European Union.
We thank Bernd Bodermann for discussions. We 
acknowledge the support of BMBF through 
project 13N13164 and of the Einstein Foundation through project ECMath OT5.

\bibliography{/home/numerik/bzfburge/texte/biblios/phcbibli,/home/numerik/bzfburge/texte/biblios/my_group,/home/numerik/bzfburge/texte/biblios/lithography,/home/numerik/bzfburge/texte/biblios/jcmwave_third_party}
%\bibliography{/home/numerik/bzfburge/texte/biblios/phcbibli,/home/numerik/bzfburge/texte/biblios/my_group,/home/numerik/bzfburge/texte/biblios/lithography,/home/numerik/bzfburge/texte/biblios/group_2012}
\bibliographystyle{spiebib}

\end{document}